\documentclass{IEEEtran}
\IEEEoverridecommandlockouts
\usepackage{cite}
\usepackage{amsthm}
\usepackage{float}
\usepackage{amsmath,amssymb,amsfonts}
\usepackage{algorithmic}
\usepackage{graphicx}
\usepackage{textcomp}
\usepackage{xcolor}
\newtheorem{theorem}{Theorem}[section]
\newtheorem{lemma}[theorem]{Lemma}
\newtheorem{definition}{Definition}[section]
\def\BibTeX{{\rm B\kern-.05em{\sc i\kern-.025em b}\kern-.08em
    T\kern-.1667em\lower.7ex\hbox{E}\kern-.125emX}}
\begin{document}

\title{A Clustering Algorithm for Correlation Quickest Hub
Discovery Mixing Time Evolution and
Random Matrix Theory}

\author{\IEEEauthorblockN{1\textsuperscript{st} Alejandro Rodríguez Domínguez}
\IEEEauthorblockA{
\textit{Miralta Finance Bank S.A., Spain}\\%Madrid, Spain \\
arodriguez@miraltabank.com}\\
\and
\IEEEauthorblockN{2\textsuperscript{nd} David Stynes}
\IEEEauthorblockA{
\textit{Munster Technological University, Ireland} \\
david.stynes@mtu.ie}
}

\maketitle

\begin{abstract}
We present a geometric version of Quickest Change Detection (QCD) and Quickest Hub Discovery (QHD) tests in correlation structures that allows us to include and combine new information with distance metrics. The topic falls within the scope of sequential, nonparametric, high-dimensional QCD and QHD, from which state-of-the-art settings developed global and local summary statistics from asymptotic Random Matrix Theory (RMT) to detect changes in random matrix law. These settings work only for uncorrelated pre-change variables. With our geometric version of the tests via clustering, we can test the hypothesis that we can improve state-of-the-art settings for QHD, by combining QCD and QHD simultaneously, as well as including information about pre-change time-evolution in correlations. We can work with correlated pre-change variables and test if the time-evolution of correlation improves performance. We prove test consistency and design test hypothesis based on clustering performance. We apply this solution to financial time series correlations. Future developments on this topic are highly relevant in finance for Risk Management, Portfolio Management, and Market Shocks Forecasting which can save billions of dollars for the global economy. We introduce the Diversification Measure Distribution (DMD) for modeling the time-evolution of correlations as a function of individual variables which consists of a Dirichlet-Multinomial distribution from a distance matrix of rolling correlations with a threshold. Finally, we are able to verify all these hypotheses.
\end{abstract}


\section{Introduction}
Financial assets portfolios are combinations of individual assets (Stocks, Bonds, Commodities, etc) that, when combined can benefit from risk diversification, and correlation structures play a crucial role in this, as seen in H. Markowitz \cite{Markowitz1952}. Therefore, knowing in advance the future behavior of correlation matrices is key for the risk management of portfolios. We focus on Quickest Change Detection (QCD) on correlation structures and Quickest Hub Discovery (QHD), variables changing their correlation. CPD and QHD allow measuring changes in the behavior of correlations to anticipate their future behavior. They also are risk management contingency measures to reallocate portfolios for better diversification and expected risk-adjusted returns when correlations change thereby improving the portfolio allocation process. Also, changes in many elements of the correlation matrix all at once are related to financial market shocks as can be seen in L. S. Junior and I. D. P. Franca \cite{Junior2011}, therefore, CPD and QHD in correlations are important too for economic and financial shock detection, which helps to preserve wealth in financial markets and the economy. 
As can be seen in A.G Tartakovsky \cite{doi:10.1080/07474940801989202}, the existing literature in CPD/QCD focuses on a setup where: 
\begin{itemize}
    \item The pre- and post- means and covariance matrices are unknown.
    \item It does not allow for high-dimensional settings where $p >> n$, p variables, and n timestamps.
    \item It assumes variables are i.i.d, a critical problem for detecting changes in correlations that assumes dependence.
\end{itemize}

As mentioned in T. Banerjee and A. Hero \cite{Banerjee2016}, to tackle this intractability, the area called sequential analysis in the literature focus on sub-optimal statistical tests with thresholds that improve performance and help with the test design. From the existing literature, the only optimal test designed for correlations and that can tackle most of the issues described, comes from work developed over the last decade by T. Banerjee et al. \cite{Banerjee2015}, T. Banerjee and A. Hero \cite{Banerjee2016}, and A. Hero and B. Rajaratnam \cite{Hero2011,Hero2012HubDI}. The objective is to detect a change in the distribution of the data matrices as quickly as possible subject to a constraint in the false alarm rate. In this line, the authors obtain an asymptotic distribution for a global and a family of local summary statistics that measures a change in the law of the Random Matrix as a way to detect the change points and isolated hubs\cite{Banerjee2016}.

The key limitation of \cite{Banerjee2015, Banerjee2016} is that they assume variables are uncorrelated before the change and correlated after. We approach the QCD/QHD test from a geometric point of view via clustering techniques. We design the geometric version of the test and prove consistency. This allows us to combine and incorporate new information into the test by defining different distance metrics. We design the test hypothesis for our geometric version via clustering performance. This allows us to compare state-of-the-art QHD test performance with a version including information about QCD to see if we can improve QHD performance. We also test the hypothesis that we can improve QHD test performance by including the pre-change time-evolution in correlation with two goals: to tackle the state-of-the-art limitation of uncorrelated pre-change variables and to see if time-evolution in correlation can improve QHD and QHD+QCD performances.

\section{Related Literature}
In S. Aminikhanghahi and D. Cook \cite{Aminikhanghahi2017}, we can see there are multiple methods and approaches for Time Series CPD. Authors mention supervised methods, such as Decision Trees, Naive Bayes, Bayesian Net, SVM, Nearest Neighbor, Hidden Markov Models, and Gaussian Mixture Models. Another group includes unsupervised methods, like Likelihood Ratio Methods(LR), Subspace Models Methods, Probabilistic Methods, Kernel-Based Methods, Graph-Based Methods, and Clustering Methods. Supervised, Probabilistic, and Clustering methods do not require extra data apart from the fitting window. Whereas LR, Subspace model, Kernel-based methods require to include post-change data\cite{Aminikhanghahi2017}. All these methods are focused on univariate time series, but our focus is on CPD in correlation structures.\par

From CPD literature, sequential analysis solutions are the most suited for our problem. QCD tries to identify the closest change point in a sequential setting. For detection to be quick, we need high-dimensional settings with $p >> n$, with n timestamps and p variables. In A. G. Tartakovsky \cite{article} we find a detailed description of sequential multi-decision for CPD and QCD. Changes in statistical properties of distributions from previously identical populations distributions are monitored to detect CPD, subject to lower levels of false alarms and delays. Methodologies rely on two standardized methods and their variants, the Page Cumulative SUM (CUSUM), and the Shiryaev-Roberts. Both had been introduced by the same author in A. G. Tartakovsky \cite{article}. The problem with this setup is that \cite{Banerjee2016}, stream independence is assumed but we want to detect changes in the level of streams’ dependence, and the setting is not high-dimensional. Alternatives that can tackle dependence in the sequential analysis literature are based on a sub-optimal test, which provides a performance analysis of the test which is then used to design the test by choosing thresholds. The efficiency of these tests is verified by simulations \cite{Banerjee2016}. However, these solutions are nonoptimal, not high-dimensional, and not focused on correlation structures. To cope with these three aspects, we need to focus on work developed by T. Banerjee and A. Hero \cite{Banerjee2015}, T. Banerjee et al. \cite{Banerjee2016}, and A. Hero and B. Rajaratnam \cite{Hero2011,Hero2012HubDI}.

As a preamble, V. Veeravalli and T.Banerjee \cite{Veeravalli2012} focused first on CPD in a parametric setting, on Bayesian CPD trying to minimize Average Detection Delay (ADD) subject to a constraint in the Probability of False Alarm (PFA). They also focused on a second solution, Minimax CPD, in which Lorden´s test from G. Lorden et al. \cite{Lorden1971} is applied. Lorden developed the first minimax theory for delays in CPD, "in which he proposed a measure of detection delay obtained by taking the supremum (over all possible change points) of a worst-case delay over all possible realizations of the observations, conditioned on the change point"\cite{Veeravalli2012}. Lorden's test is important to understand their posterior work.

In \cite{Hero2011}, a discovery is a correlation above a threshold, they derive an asymptotic expression for the mean number of discoveries that is a function of the number of samples. It is shown that the mean number of discoveries is influenced by the population covariance matrix by the Bhattacharyya measure of the average pairwise dependency of the p-multivariate U-scores defined on the (n-2)-dimensional hypersphere \cite{Hero2011}. Under weak dependency assumptions, the number of discoveries is asymptotically represented by a Poisson distribution. For auto-correlation and cross-correlation discoveries this Poisson distribution is measured by the number of positive vertex degrees in the associated sample correlation graph \cite{Hero2011}. In \cite{Hero2012HubDI}, they focus on hub discovery in partial correlation graphs, and an extension for variables with a specific degree of connectivity. A hub is defined broadly as any variable that is correlated with at least $\delta$ other variables having a magnitude correlation exceeding $\rho$. Their setup is the first high-dimensional of its kind. They show that the count $N_{\delta,\rho_p}$ of the number of groups of $\delta$ mutually coincident edges in the correlation graph (and partial) with correlation threshold $\rho$ converges to a Poisson variable:

\begin{equation}
 P(N_{\delta,\rho_p}>0)\rightarrow \exp(-\Lambda/\varphi(\delta))   
\end{equation}

This had implications for future work developed in \cite{Banerjee2015, Banerjee2016}. In \cite{Banerjee2015}, authors introduce a nonparametric QCD test for large-scale random matrices based on a global summary statistic from asymptotic properties of RMT. It is assumed pre- and post- change distributions of the i.i.d random matrices rows are unknown or belong to an elliptically contoured family \cite{Banerjee2015}. If pre- and post- change densities  $f_X^0$ and $f_X^1$ are known, and the mean $\mu_m$ is constant before and after the change, algorithms such as Cumulative Sum (CumSum) or Shiryaev-Roberts (SR) can be efficiently used as both have optimal properties respect to Lorden formulations. In this case, it is a parametric CPD problem, and asymptotically optimally solved by Generalized Likelihood Ratio (GLR) tests. For \cite{Banerjee2015}, pre- and post- change densities are unknown and present an optimal nonparametric solution, as an asymptotically optimal solution to the minimax CPD in the random matrix setup using large-scale RMT. The framework in \cite{Banerjee2015} is suited for high-dimensional settings. Therefore, a summary to justify why we focus on nonparametric methods with RMT like \cite{Banerjee2015, Banerjee2016}: 
\begin{itemize}
    \item Pre- and post- change densities  $f_X^0$ and $f_X^1$ are NOT known, and $\mu_m$ is NOT constant before and after the change.
    \item Need to focus on high-dimensional settings $p > > n$ for QCD and QHD in correlation structures.
    \item Other CPD/QHD methods are rejected too because cannot tackle dependence.
\end{itemize}

\section{Framework Description}

We now explain the approach in \cite{Banerjee2015, Banerjee2016}, how the densities of the summary statistics are obtained and used to test for QCD and QHD. 
\subsection{QCD}
Given an elliptically distributed random data matrix $\mathbb{X}$, such that,  $\mathbb{X}=\left[X_1,\dots,X_p\right]=\left[X_{(1)}^T,\dots,X_{(n)}^T\right]^T$.\par
$X_i=\left[X_{1i},\dots,X_{ni}\right]^T$, is the $i^{th}$ column and, $X_{(i)}=\left[X_{i1},\dots,X_{ip}\right]$, is the $i^{th}$ row. The sample covariance matrix is:

\begin{align}
    S\ =\ \frac{1}{n-1}\sum_{i=1}^{n}{{(X_{(i)}\ -\ \bar{X})}^T(X_{(i)}\ -\ \bar{X})}
\end{align}
$\bar{X}$ is the sample mean of the n rows of $\mathbb{X}$. The sample correlation matrix is:
\begin{align}
    R\ =\ D_S^{-1/2}S{\ D}_S^{-1/2}
\end{align}
with $D_S$ the diagonal matrix of S. $R_{ij}$, the element in the $i^{th}$ row and $j^{th}$ column of the matrix R, is the sample correlation coefficient between the $i^{th}$ and $j^{th}$ columns of $\mathbb{X}$. 

Then, $d_{NN}^{(k)}(i)$ is the k-nearest neighbor of the $i^{th}$ column of $\mathbb{X}$ in correlation distance: $d_{NN}^{(k)}(i) = k^{th}$ largest order statistic of $\left\{\left|R_{ij}\right|;j\neq i\right\}$. In other words, is the k-nearest neighbor in correlation distance of the variable $i^{th}$ from the data matrix $\mathbb{X}$, with $X_i=\left[X_{1i},...,X_{ni}\right]^T$, and n samples.

Then the summary statistic is, for a fixed k \cite{Banerjee2015}: 
\begin{align}V_k(\mathbb{X})=\max_{i}{d_{NN}^{(k)}(i)}\end{align}

The distribution of the statistic $V_k$ can be related to the distribution of an integer value random variable $N_{\delta,\rho}$. This number is the total number of hubs in the correlation graph $\mathcal{G}_p(R)$, a hub being a vertex with degree $\delta$, if $\delta_i\geq\delta$ for vertex i with degree $\delta_i$ in the correlation graph $\mathcal{G}_p(R)$. For a threshold correlation $\rho\in\left[0,1\right]$, $\mathcal{G}_p(R)$ is the correlation graph associated with correlation matrix R, as an undirected graph with p vertices, each being the columns of data matrix  $\mathbb{X}$. An edge is present between vertices i and j if the magnitude of the sample correlation coefficient between the $i^{th}$ and $j^{th}$ components of the random vector X is greater than $\rho$, $\left|R_{ij}\right|\geq\rho,\ i\neq j$ \cite{Banerjee2015}. $N_{\delta,\rho}$ is the total number of hubs in the correlation graph, $N_{\delta,\rho}=card\left\{i:\delta_i\geq\delta\right\}$. The events, ${V_\rho(X)\geq\rho}$, and, ${N_{\delta,\rho}>0}$, are equivalent. And so we have \cite{Banerjee2015}: 

\begin{align}  P(V_\rho(X)\geq\rho)=P(N_{\delta,\rho}>0)
\label{equation5}
\end{align}

In a high-dimensional setting $p\rightarrow\infty$ for n fixed, $P(N_{\delta,\rho}>0)$ is asymptotically approximated by relating  $N_{\delta,\rho}$ to a Poisson distribution. The proof is in \cite{Hero2012HubDI}. Using this theorem, the large p distribution of $V_k$ defined in (\ref{equation5}) can be approximated, for $k=\delta$, by:

\begin{align}P(V_\delta(X)\le\rho)=\ \exp(-\ \Lambda(\rho)\ J_{X}/\phi(\delta)),\ \rho\in\left[0,1\right]\end{align}

For $\rho>0$ and large p, $V_{\delta}$ has density:

\begin{align}
f_V(\rho)=\ -\ \frac{\Lambda^{'}(\rho)}{\phi(\delta)}\ J_X\ \exp(-\ \frac{\Lambda(\rho)}{\phi(\delta)}J_X),\ \rho\in(0,1]
\end{align}

If $\delta=1$ fixed, $V_{\delta}$ reduces to the nearest neighbor (correlation) distance: \begin{align}V(X)=\max_{i\neq j}{\left|R_{ij}\right|}\end{align} and the density reduces to: 

\begin{align}f_V(\rho;J)=\ \frac{C}{2}{(1-\rho^2)}^\frac{n-4}{2}\ J\ \exp(-\ \frac{C}{2}J T(\rho)),\ \rho\in(0,1] 
\label{equation10}
\end{align}

The QCD test is performed by computing the sequence of global summary statistics $\left\{V_\delta(\mathbb{X}(m))\right\}_{m\geq1}$ from the data matrix sequence. $J_0$ and $J_1$ are the values of J before and after the change. If the pre-change dispersion matrix $\Sigma_{0}$ is diagonal, $J_{0}=1$. The QCD test takes the form, defined by stopping time $\tau_{G}$:
\begin{align}\tau_{G}=\inf_{m\geq1}{\left\{\max_{1\le l\le m}\sup_{J:\left|J-1\right|\geq\epsilon}\sum_{i=l}^{m}{\log\frac{f_V(V(i);J)}{f_V(V(i);1)}}>A\right\}}\end{align}

The parameter A is a threshold to control for the false alarm rate, $\epsilon$ is the minimum magnitude of change in correlation. 

\subsection{QHD}

In \cite{Banerjee2016}, for a nondiagonal covariance matrix $\Sigma$ with correlation coefficients ${\rho_{ki}}$:
\begin{align}
    V_k(\Sigma)=\max_{i\neq k}{\left|\rho_{ki}\right|},\ \ \ k\in\left\{1,\dots,p\right\}\
\end{align}
is the maximum magnitude correlation coefficient for the $k^{th}$ variable. Hubs are defined as:

\begin{align}
H\ =\ \left\{k\colon \ V_k=\ \max_{1\le j\le p}{V_j}\right\}\ 
\end{align}

The local summary statistics are defined:
\begin{align}
  V_k(\mathbb{X})=\ V_k(R)\ =\ \max_{i\neq k}{\left|R_{ki}\right|,\ \ \ k\in\left\{1,\dots,p\right\}}  
\end{align}

The large p distribution of $V_k$ can be approximated:

%\begin{IEEEeqnarray}{R}
%P(V_k(X)\le\rho)\sim \exp(-\ \Lambda_{k,\rho})\ %=\exp(-(p-1)P_0(\rho)J_k),\IEEEnonumber\\ \rho\in\left[0,1\right]   
%\end{IEEEeqnarray}

\begin{align}
\begin{split}
P(V_k(X)\le\rho)\sim \exp(-\ \Lambda_{k,\rho})=\ \exp(-(p-1)P_0(\rho)J_k),\\ \rho\in\left[0,1\right]    
\end{split}
\end{align}

For each k, the density $f_{V_k}$ is a member of a one-parameter exponential family with $J_k$ as the unknown parameter. We have for $\rho\in\ \left[0,1\right]$, the exponential family form of the density $f_V$ with parameter $J_k$: 

\begin{align}f_V(\rho;J_k)=\frac{2(p-1)J_k{(1-\rho^2)}^\frac{n-4}{2}}{B((n-2)/2,1/2)}\exp(-(p-1)J_k P_0(\rho))
\label{equation19}
\end{align}

the variable V tends to take on higher values as the parameter $J_k$ increases \cite{Banerjee2016}. The QHD is reduced to the following GLR-based family of QCD tests:

\begin{align}
    \tau_V=\min_k{\tau_V^{(k)}}=\min\left\{m:\ \max_{1<k<p}{G_k(m)>A_v}\right\}
\end{align}

where:
\begin{align}
G_k(m)=\max_{1\le l\le m}{\sup_{J_k:\left|J_k-1\right|\geq\epsilon_v}{\sum_{i=l}^{m}\log\left(\frac{f_V\left(V_k\left(i\right);J_k\right)}{f_V\left(V_k\left(i\right);1\right)}\right)}}
\label{equation20}
\end{align}

the test $\tau_V$ can be used to detect if the parameter of any of the p variables is affected by the change \cite{Banerjee2016}. $\epsilon_V$ represents the minimum magnitude of change, away from $J_k = 1$, that the user wishes to detect.$A_v$ is the threshold to control for false alarms and delays. Finally, for the joint detection and hub discovery, the authors combine local (QHD) and global (QCD) tests solutions with the maximum stopping times:
\begin{align}
\tau_{HB}=\max\left\{\tau_V,\tau_U\right\}
\label{equation18}
\end{align}

If, $D_k(\mathbb{X}(1),\dots,\mathbb{X}(\tau_{HB}))$, is the binary decision variable with 1 if k is a hub. The rule for hub discovery is \cite{Banerjee2016}, for a fix positive integer q:
\begin{align}
    D_k(\mathbb{X}(1),\dots,\mathbb{X}(\tau_{HB}))=\ \mathbb{I}_{\left\{G_k(\tau_{HB})\ is\ top\ q\ statistic \right\}}
\end{align}

\begin{figure}
\vspace{-23mm}
\setlength\abovecaptionskip{-7\baselineskip}
	\centering
	\includegraphics[width=80mm]{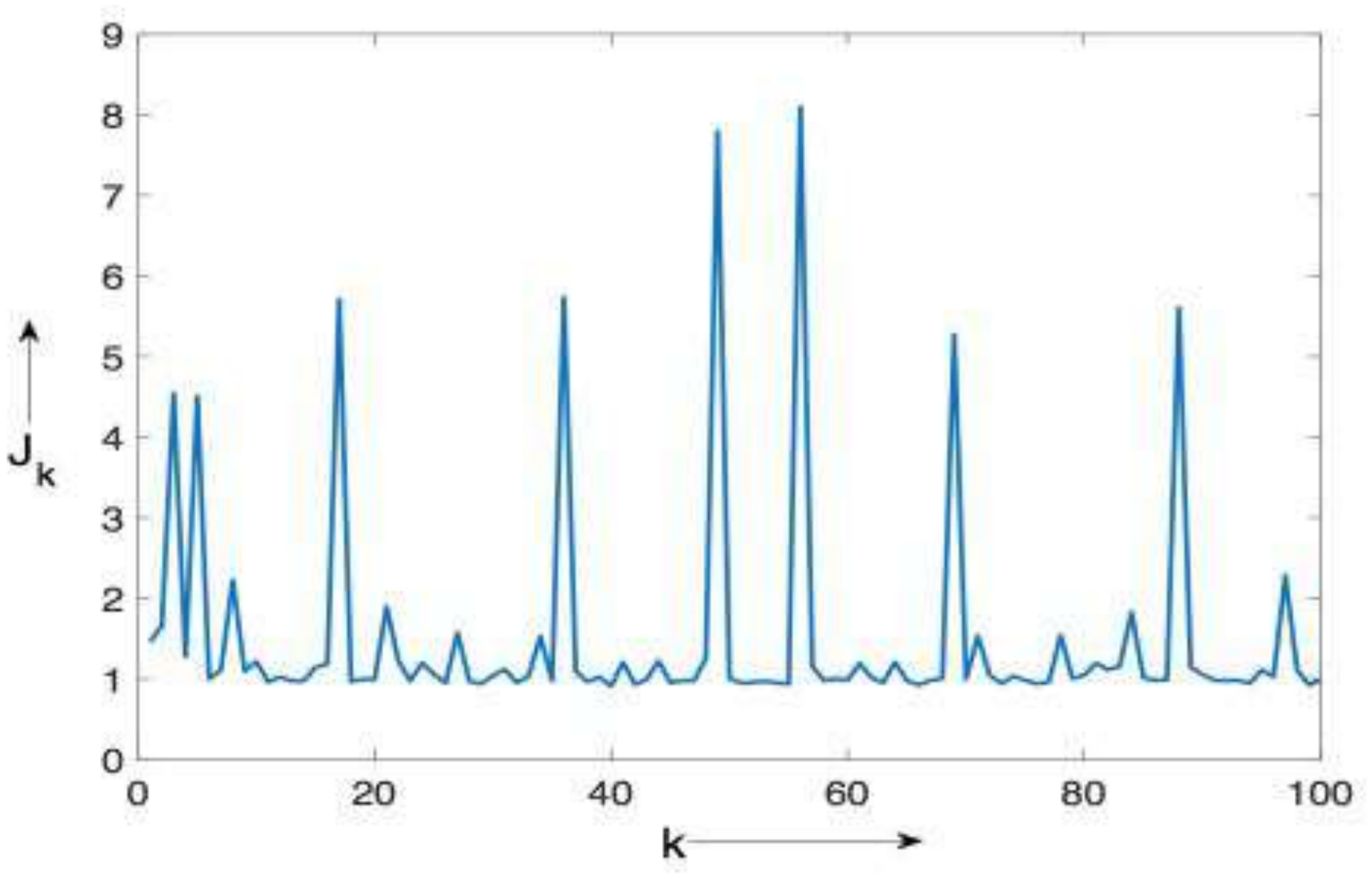}
	\caption{$J_k$ values for 100 variables. Adapted from \cite{Banerjee2016}}
	\label{figure2}
\end{figure}

\begin{figure}
\vspace{-23mm}
\setlength\abovecaptionskip{-7\baselineskip}
	\centering
	\includegraphics[width=80mm]{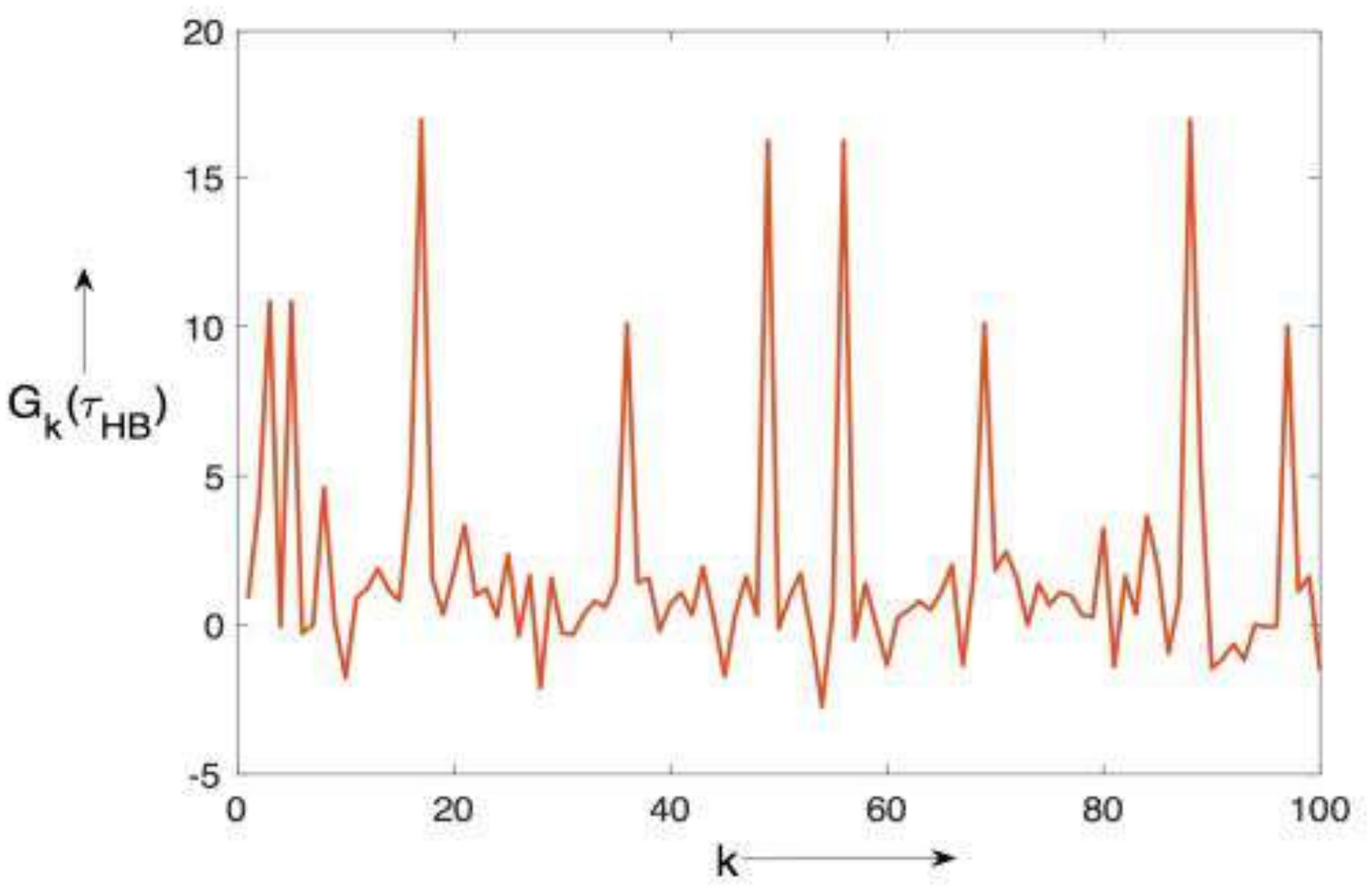}
	\caption{$G_k$ values at the time of stopping $\tau_{HB}$ for 100 variables. Adapted from \cite{Banerjee2016}}
	\label{figure3}
\end{figure}

\subsection{Geometric QCD and QHD via clustering}
In this section, we focus on our contribution, which consists of a geometric version of the tests via clustering. For that, we prove consistency and design test hypotheses. 
\begin{lemma}
The times with maximum parameter J and maximum density of the global summary statistic converges in probability, and more if J is closer to its maximum. These times coincide in probability with the optimal solution for QCD in correlation structures:

\begin{align}
\begin{split}
\left(\max_i{\left(J(i)\right)}{\xrightarrow p}\max_i{f_V(V(i);J)}\right)|\ J\rightarrow\ \max_J{(J(i))},\\ \forall\ i=1,\dots,m
\end{split}
\end{align}
\label{lemma1}
\end{lemma}
\begin{IEEEproof}
The proof can be found in Appendix A
\end{IEEEproof}

The same applies for QHD test for both, $J_k$ and $G_k$. Clustering variables based on the value of $J_k$ and $G_k$, or equivalently maximum density values ($f_V(\rho;J_k)$, is statistically consistent with tests in \cite{Banerjee2015,Banerjee2016}. From Figures \ref{figure2} and \ref{figure3}, we can see that "what is clear is that high $G_k$ values at the time of stopping correspond to high $J_k$ values of the summary statistics, and these values together have significantly higher magnitude than the values for the rest of the variables. This motivates the pick-the-top approach. In fact, in practice, one can plot the $G_k$ values this way and identify all the variables that have experienced a change in correlation" as stated in \cite{Banerjee2016}.

\begin{lemma}
The rule for joint detection and hub discovery in the geometric test is given by the following decision variable:
\begin{align}
\begin{split}
 D_k(\mathbb{X}(1),\dots,\mathbb{X}(\tau_{HB}))=\mathbb{I}_{\left\{k\in S_j(\tau_{HB})\right\}},\\
\left\{k:{\left\{G_k(\tau_{HB})\ is\ top\ q\ statistic \right\}};j=1,\dots,w\right\}\subset\\ 
\subset S_j(\tau_{HB})
\end{split}   
\end{align}
with change point given in this case by: 

\begin{align}
    \tau_{HB}=\left\{\tau_U\vee\ \tau_V\right\}   
\end{align}
\label{lemma2}
\end{lemma}

\begin{IEEEproof}We apply lemma \ref{lemma1} to $G_k$:
\begin{align}
\begin{split}
\forall\ i_1 \neq i_2\ \ \exists\ k :\\ 
\left[\left(\tau_v{\xrightarrow p}\left[\max_k{\left(G_k(i_2)\right)}\right]\right)\nLeftrightarrow
\left(\tau_v{\xrightarrow p}\left[\max_k{\left(G_k(i_1)\right)}\right]\right)\right]\\
\Rightarrow
\left[i_1=i_2=i,\ i\in \{\tau_v,\tau_U,t\}\right],\\ \left[\{t\neq\tau_v\}\ \land \{t\neq\tau_U\}:\ i=\{\tau_v\ \vee\tau_U\vee t\}\right]\\
\Rightarrow\left[\tau_{HB}=i\ : \ i=\{\tau_v\ \vee\tau_U\}\right]\ 
\end{split}
\end{align}
\end{IEEEproof}
\begin{lemma}
Geometric test consistency via clustering:
\begin{align}
\begin{split}
A\ \ =\ \ \frac{\frac{1}{nT}\sum_{k=1}^{n}\sum_{m=1}^{T}{D_k(\mathbb{X}(1),\dots,\mathbb{X}(\tau_{HB}=m))\ \ }}{\frac{1}{nT}\sum_{k=1}^{n}\sum_{m=1}^{T}{{\mathbb{I}\ }_{\left\{k\in S_j(HB);\ \tau_{HB}=m\right\}}}},\\
A \xrightarrow p\ 1\ \ as \ \ T\rightarrow\infty\ ; \ A\ \ {\rightarrow\ 1} \ as\ \ T, G_k, J_k\rightarrow\infty,\\
\forall\ k=1,\dots,n, \forall\ S_j\ ,\ j=1,\dots,w,\\
\tau_{HB}=\left\{\tau_U (m)\vee\ \tau_V (m)\right\}, m=1,\dots,T
\end{split}
\end{align} 
\label{lemma3}
\end{lemma}
\begin{IEEEproof}We apply lemmas \ref{lemma1} and \ref{lemma2}\ and the Law of Large Numbers
\end{IEEEproof}
\begin{lemma}
Test hypothesis for geometric QHD comparison is based on clustering performance.
\end{lemma}
\begin{IEEEproof}
\begin{align}
\begin{split}
\forall\ \left\{a_1,\dots,a_p\right\}\in H_m,\\
CPL\equiv \left\{j:\ \left\{a_s:\ G_s(\tau_{HB})\ is\ top\ 1\ statistic\ \right\}\subset S_j\right\},\\
\forall\ m=1,\dots,T,\{ \forall\ \tau_{HB}|
\left\{\tau_{HB}=\left\{\tau_U\vee\ \tau_V\right\},
\tau_{HB}\in m\right\}\},\\
\forall\ S_j\ ,\ j=1,\dots,w, \forall\ \ p\le n:\\
LPM (\tau_{HB})=\ \frac{\sum_{i=1}^{p}\mathbb{I}_{\left\{\ a_i \subset S_j\ |\ j=CPL \right\}}}{p}\\
C(\tau_{HB})=\sum_{i=1}^{m}\mathbb{I}_{\left\{\tau_{HB}\equiv i \right\}}\\
MCP (\left\{1,\dots,m\right\})=\frac{1}{C(\tau_{HB})}\sum_{i=1}^{C(\tau_{HB})}{LPM(\tau_{HB}(i))} 
\end{split}
\end{align}
Local Performance Metric (LPM) is the clustering performance at $\tau_{HB}$. Mean Clustering Performance (MCP) is the average performance over all $\tau_{HB}$ in a sample with n variables, m timestamps, and hubs $H_m$ of p variables.
\end{IEEEproof}
\begin{definition}
The QCD + QHD distance metric between two variables X and Y at a time m is given by:
\begin{align}
\begin{split}
{(QCD+QHD)}_{ij}(m)=F\left({QHD}_{ij}(m),g_U(U(m);J),\theta\right)=\\
=\theta_1{\rm QHD}_{ij}(m)+\\
+\theta_2\frac{\left|\max_k{f_V(V_i(k);J_i)}-\max_k{f_V(V_j(k);J_j)}\right|}{\max_k{g_U(U(k);J_k)}},\\ 
\forall\ i,j = 1,\dots,n,\ i\neq j,\ k=1,\dots,m;\ m=1,\dots,T, \\  \theta_1+\theta_2=1 
\end{split}
\end{align}
\end{definition}
The Euclidean Distance between two rolling correlations is defined as:  

\begin{align}
\begin{split}
 TE_{ij}(k)=\sqrt{\sum_{k=t-w}^{k}{({rc}_{ik}-{rc}_{jk})}^2\ }\\ \ \forall\ i,j=(p,m)\ |p\neq m,\ \ i\neq j\   
\label{equation26}
\end{split}
\end{align}

with k being the timestamp when the distance is computed, based on previous rolling correlation time series. i and j the tuples of variables p and m.

\begin{definition} The Diversification Measure Distribution (DMD) is a node degree distribution of a composite network. The first subnetwork is an unweighted network of distance labels from a filtered distance matrix of rolling correlations with a closeness threshold. Nodes are the rolling correlations pair variables and edge the distances labels. The second subnetwork (the asset variables network) is an unweighted network of rolling correlations labels that are the result of the first subnetwork, with nodes the individual assets and edges the labels of the rolling correlations.
\begin{lemma}The link between both subnetworks is given by the Bayes Theorem. The conjugacy by Bayes Theorem shows, as a result, the Dirichlet-Multinomial distribution, a compound distribution which is the degree distribution of the compounded network, from two distributions which are the degree distributions of each subnetwork (a subnetwork of rolling correlations through time with the Categorical distribution and the subnetwork of asset variables with the Dirichlet distribution).
\label{lemma5}
\end{lemma}
\begin{IEEEproof} The proof can be found in Appendix B. \end{IEEEproof}
\begin{align}
\begin{split}
P\left(\theta\ \middle| D\right) \ \sim\ P\left(D\middle|\theta\right)P\left(\theta\right)=\ \prod_{i=1}^{n}\prod_{k=1}^{m}{\theta_k^{rc_k+\ a_k-1}}\sim \\
\sim Dir\left(\theta\ \middle| rc+a\right)=Dir\left(d\le \Phi\ \middle| rc+a\right),\\
rc_k=\ I\left[{rc}_i=k_1\right]+I\left[{rc}_i=k_2\right],\\  
a_k=I\left[a_i=k_{11}\right]+I\left[a_i=k_{12}\right]+I\left[a_i=k_{21}\right]+I\left[a_i=k_{22}\right]
\end{split}
\label{equation30}
\end{align}
\end{definition}

where $a_i$, is the asset variable i, d is the distance TE from (\ref{equation26}) on each iteration, n is the number of asset variables in the sample, m is the number of correlations associated with the variables dataset. $k_{11},\dots,k_{22}$ are the different names of the variables that are part of the two pairs of rolling correlations for distances $d=TE_{k_1k_2}(k)$, with $k_1$ and $k_2$ correlation labels (pair of variables). Equation (\ref{equation30}) is the distribution that counts the number of times $a_i$ appears in any of $k_{11},\dots,k_{22}$, for distances $d=TE_{k_1k_2}(k)$ below a closeness threshold $\Phi$, or rolling correlations with similar time-evolution.

\begin{definition}
The similarity metric in terms of time-evolution of correlation is given by the difference in DMD:

\begin{align}
\begin{split}
DD_{ij}(m)=\ \left(P(d\le \Phi|a_i, rc_k)\ -\ P(d\le \Phi|a_j, rc_p)\right)=\\ \left[\left(\prod_{i=1}^{n}\prod_{k=1}^{m}{\theta_k^{rc_k+\ a_k-1}}\right) - \left(\prod_{j=1}^{n}\prod_{p=1}^{m}{\theta_p^{rc_p+\ a_p-1}}\right)\right],\\
\forall\ i,j = 1,\dots,N,\ i\neq j,\ k=1,\dots,M,\ p =1,\dots,M,\\ k\neq p
\label{equation31}
\end{split}
\end{align}

\end{definition}

\begin{figure}
\vspace{-22mm}
\setlength\abovecaptionskip{-9\baselineskip}
	\centering
	\includegraphics[width=87mm]{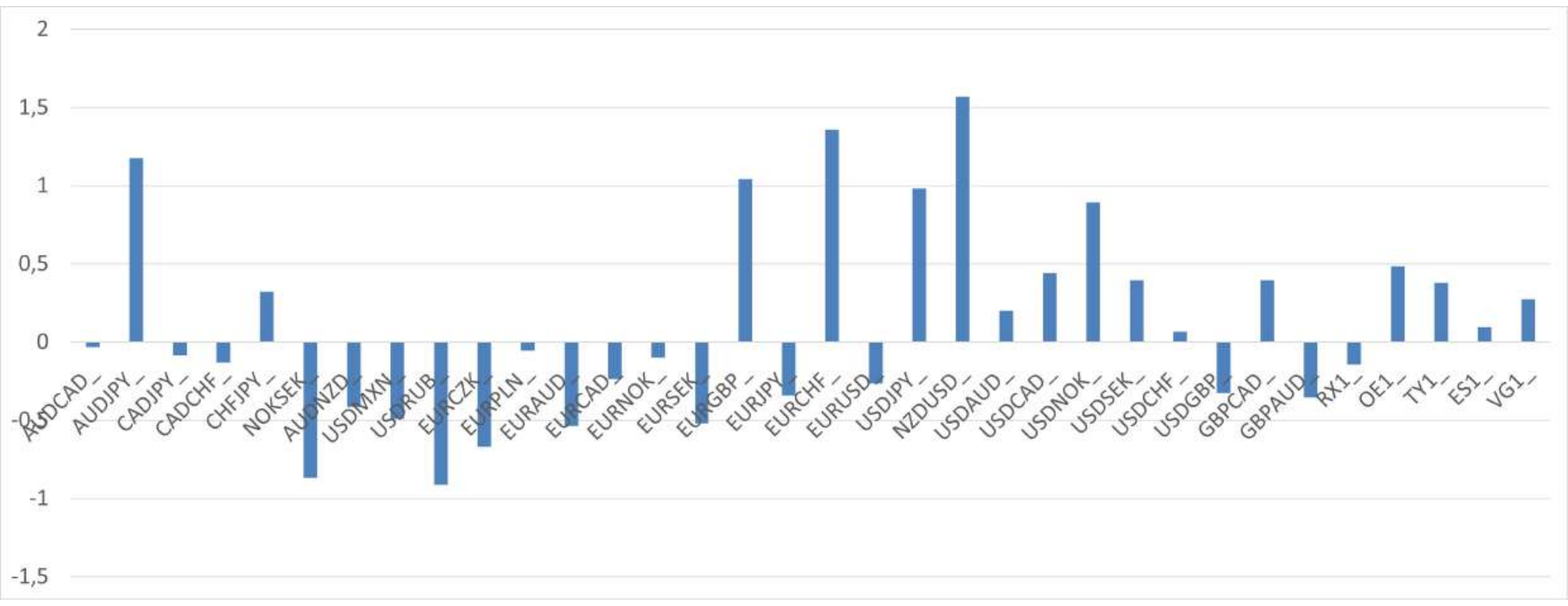}
	\caption{DMD for a particular timestamp. DMD in vertical axis and financial assets in horizontal axis}
	\label{figure31}
\end{figure}

\begin{figure}
\vspace{-22mm}
\setlength\abovecaptionskip
{-7\baselineskip}
	\centering
    \includegraphics[width=87mm]{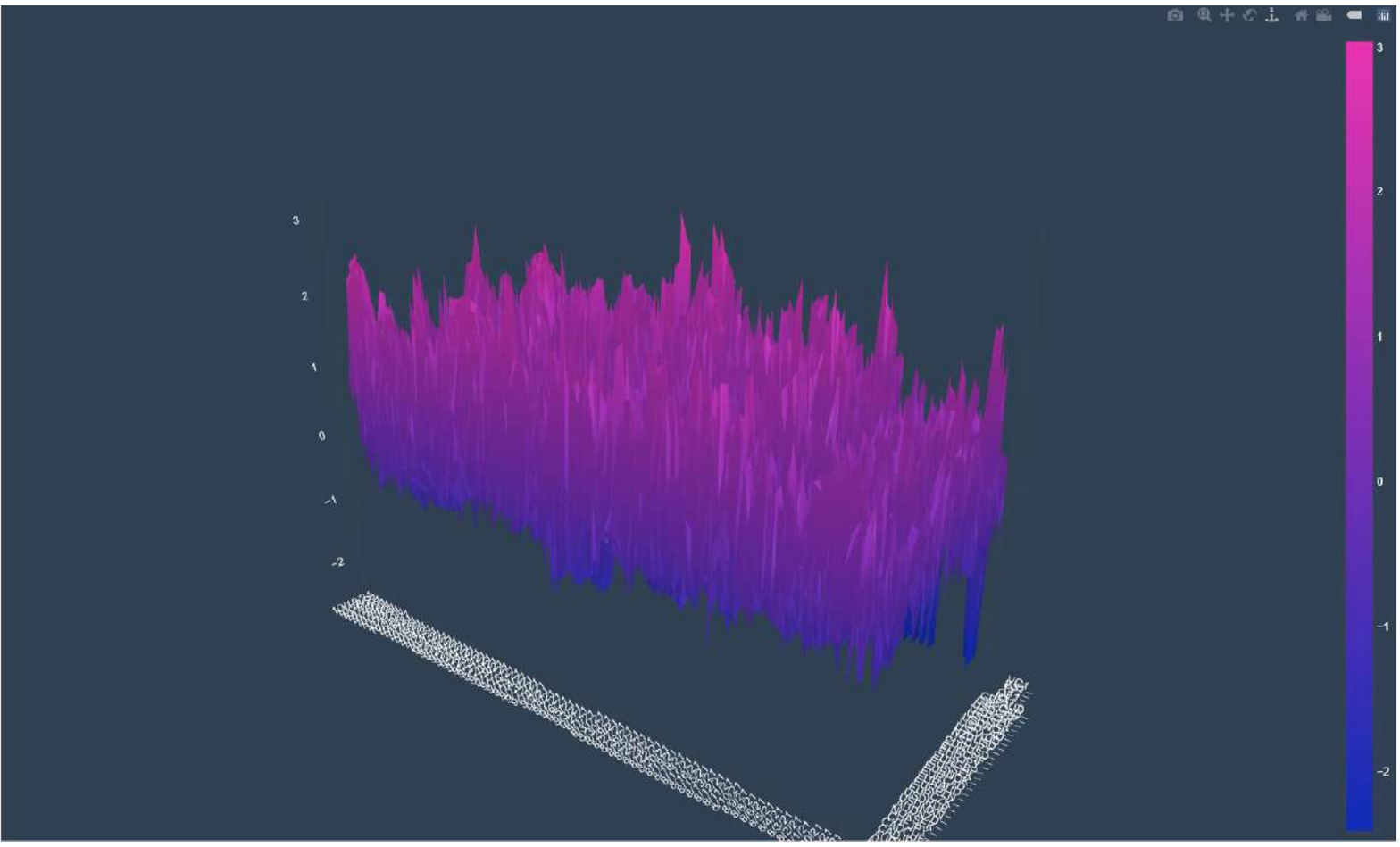}
	\caption{DMD for a sample of 1500 timestamps. The two horizontal axis are the categories (financial assets;right axis) and the timestamps with dates (hourly time series;left axis)}
	\label{figure32}
\end{figure}

We standardize the distribution with respect to the sample of 1500 timestamps and show the distribution for a particular timestamp in Figure \ref{figure31}, with assets in the horizontal axis. The higher the bar for an asset, the less diversified is in terms of time-evolution of correlation. We show the distribution in time for the 1500 timestamps in Figure \ref{figure32}. The Test Hypothesis distance metric is given by:

\begin{align}
\begin{split}
 TH_{ij}(k)=G\left({(QCD+QHD)}_{ij}(k),DD_{ij}(k),\gamma\right)=\\
 \gamma_1{(QCD+QHD)}_{ij}(k)+\gamma_2 DD_{ij}(k),\\   
 \forall\ i,j = 1,\dots,N,\ i\neq j,\ k=1,\dots,M,\\
 \gamma_1+\gamma_2=1, \theta_1+\theta_2=\gamma_1  
\end{split}
\label{equationTH}
\end{align}

Given a set of variables k, $k=\left\{1,...,N\right\}$, candidates solutions for QHD in correlations, we aim to partition the k into $w(\neq k)$ sets $S=\left\{S_1,..,S_{w\ }\right\}$ so as to minimize the within-cluster sum of squares. The objective is, for a sample of time stamps m, $m=\left\{1,...,T\right\}$, to run sequentially the following algorithm:
\begin{align}
\begin{split}
\min_{S(m)}{\sum_{j=1}^{w}\frac{1}{\left|S_j(m)\right|}\sum_{p=1}^{T}\sum_{k_1, k_2\ \in\ S_j(m)}{{(TH)}_{k_1, k_2}(m)}},\\
\forall\ k_1,k_2\in k;\ k_1\neq k_2
|\left[\left({\ k}_1\in x_1\right)\vee\ \left({\ k}_2\in x_1\right)\right]\land\\
\land\left[\left({\ k}_1\in x_2\right)\vee\ \left({\ k}_2\in x_2\right)\right]    
\end{split}
\end{align}

\section{Numerical Results}
Test hypotheses are carried on financial asset correlations hourly time series from a dataset of 38 assets of different classes and 3500 hourly prices. Tables \ref{Table1} and \ref{Table2} show uni-sample test hypotheses for different model configurations. We compare QHD, QHD+QCD, and QHD+QCD+DD, with labels 1, 2, and 3 respectively for the Tables. Each row represents a different model configuration, with an array of parameters (second column). We compute the mean (MCP), minimum, and standard deviation of the LPM for a 100-hours sample. In Table \ref{Table3}, we show test hypotheses from multiple 100-hours samples. Model configuration is given as: $J_k$, threshold $A_k$, $\theta$ and $\gamma$ the distance metric parameters in (\ref{equationTH}), K number of clusters, $\Phi$ the time-evolution threshold, in this order.\par

\begin{table}[H]
\caption{Uni-sample Test Hypotheses. Uncut dataset}
\centering
\begin{tabular}{c|c|c|c|c}
\hline
 Label & $J_k$\ \ $A_k$\ \ $\theta$\ \ $\gamma$\ \ K\ \ $\Phi$ & Mean     & Min   & Std  \\
  \hline\hline
1 & $2\ \ 0\quad \ 1\quad \ 0\quad \ \ 5\ \ 2$            & 75,00\%  & 44\%  & 21\% \\
\hline
3 & 2\ \ 0\ \ 0.95\ \ 0.10\ \ 5\ \ 2      & 81,94\%  & 44\%  & 21\% \\
\hline
3 & 2\ \ 0\ \ 0.98\ \ 0.05\ \ 5\ \ 2       & 83,80\%  & 39\%  & 22\% \\
\hline
3 & 2\ \ 0\ \ 0.98\ \ 0.05\ \ 5\ \ 4       & 92,67\%  & 63\%  & 15\% \\
\hline
1 & $2\ \ 3\quad \ 1\quad \ 0\quad \ \ 5\ \ 2$            & 89,23\%  & 46\%  & 22\% \\
\hline
3 & 2\ \ 3\ \ 0.98\ \ 0.05\ \ 5\ \ 4       & 96,59\%  & 50\%  & 11\% \\
\hline
1 & $5\ \ 0\quad \ 1\quad \ 0\quad \ \ 5\ \ 2$            & 74,39\%  & 43\%  & 24\% \\
\hline
2 & 5\ \ 0\ \ 0.02\ \ \ 0\ \ \ \ \ 5\ \ 2       & 87,83\%  & 33\%  & 23\% \\
\hline
1 & $5\ \ 3\quad \ 1\quad \ 0\quad \ \ 5\ \ 2$            & 92,22\%  & 30\%  & 22\% \\
\hline
2 & 5\ \ 3\ \ 0.10\ \ \ 0\ \ \ \ \ 5\ \ 2        & 92,59\%  & 33\%  & 21\% \\
\hline
3 & 5\ \ 3\ \ 0.02\ \ 0.05\ \ 5\ \ 2       & 96,22\%  & 36\%  & 15\% \\
\hline
3 & 5\ \ 3\ \ 0.95\ \ 0.05\ \ 5\ \ 4       & 97,58\%  & 78\%  & 7\%  \\
\hline
1 & $10\ \ 3\quad \ 1\quad \ 0\quad \ \ 5\ \ 2\ $            & 92,31\%  & 62\%  & 15\% \\
\hline
2 & 10\ \ 3\ \ \ 0.50\ \ 0\ \ \ \ 5\ \ 4\ \      & 93,55\%  & 68\%  & 13\% \\
\hline
3 & 10\ \ 3\ \ 0.95\ \ 0.05\ \  5\ \ 4\  \     & 95,38\%  & 31\%  & 17\% \\
\hline
1 & $100\ \ 0\quad \ 1\quad \ 0\quad \ \ 5\ \ 2\ \ \ $            & 85,38\%  & 67\%  & 13\% \\
\hline
3 & 100\ \ 0\ \ 0.98\ \ 0.10\ \  5\ \ 2\  \ \ \    & 85,94\%  & 43\%  & 22\% \\
\hline
1 & $100\ \ 3\quad \ 1\quad \ 0\quad \ \ 5\ \ 2\ \ \ $            & 98,08\%  & 92\%  & 3\%  \\
\hline
1 & $1000\ \ 0\quad \ 1\quad \ 0\quad \ \ 5\ \ 2\ \ \ \ $            & 97,44\%  & 92\%  & 4\%  \\
\hline
1 & $1000\ \ 3\quad \ 1\quad \ 0\quad \ \ 5\ \ 2\ \ \ \ $            & 100,00\% & 100\% & 0\% \\

\end{tabular}

\label{Table1}
\end{table}

\begin{table}[H]
\caption{Uni-sample Test Hypotheses. Cut dataset}
\centering
\begin{tabular}{c|c|c|c|c}
\hline
 Label & $J_k$\ \ $A_k$\ \ $\theta$\ \ $\gamma$\ \ K\ \ $\Phi$ & Mean     & Min   & Std  \\
  \hline\hline
1 & $2\ \ 0\quad \ 1\quad \ 0\quad \ \ 5\ \ 2$            & 85,71\%  & 50\%  & 23\% \\
\hline
2 & 2\ \ 0\ \ \ 0.98\ \ 0\ \ \ \ 5\ \ 2       & 91,87\%  & 59\%  & 16\% \\
\hline
3 & 2\ \ 0\ \ 0.50\ \ 0.15\ \ 5\ \ 4       & 92,53\%  & 55\%  & 17\% \\
\hline
1 & $2\ \ 3\quad \ 1\quad \ 0\quad \ \ 5\ \ 2$            & 91,95\%  & 52\%  & 18\% \\
\hline
3 & 2\ \ 3\ \ 0.98\ \ 0.05\ \ 5\ \ 2       & 92,26\%  & 61\%  & 15\% \\
\hline
3 & 2\ \ 3\ \ 0.50\ \ 0.15\ \ 5\ \ 4       & 94,25\%  & 66\%  & 13\% \\
\hline
1 & $5\ \ 0\quad \ 1\quad \ 0\quad \ \ 5\ \ 2$            & 88,51\%  & 31\%  & 26\% \\
\hline
2 & 5\ \ 0\ \ \ 0.95\  0\ \ \ \ \ 5\ \ 2       & 89,10\%  & 46\%  & 20\% \\
\hline
3 & 5\ \ 0\ \ 0.98\ \ 0.05\ \ 5\ \ 2       & 90,21\%  & 50\%  & 19\% \\
\hline
3 & 5\ \ 0\ \ 0.95\ \ 0.05\ \ 5\ \ 2       & 91,67\%  & 33\%  & 22\% \\
\hline
1 & $5\ \ 3\quad \ 1\quad \ 0\quad \ \ 5\ \ 2$            & 95,40\%  & 72\%  & 10\% \\
\hline
2 & 5\ \ 3\ \ 0.95\ \ 0\ \ \ \ \ 5\ \ 2       & 96,25\%  & 50\%  & 11\% \\
\hline
3 & 5\ \ 3\ \ 0.98\ \ 0.05\ \ 5\ \ 2       & 96,30\%  & 70\%  & 10\% \\
\hline
1 & $10\ \ 0\quad \ 1\quad \ 0\quad \ \ 5\ \ 2\ $            & 90,80\%  & 45\%  & 21\% \\
\hline
1 & $10\ \ 3\quad \ 1\quad \ 0\quad \ \ 5\ \ 2\ $            & 94,83\%  & 69\%  & 12\% \\
\hline
3 & 10\ \ 3\ \ 0.95\ \ 0.02\ \ 5\ \ 4\  \     & 95,00\%  & 55\%  & 14\% \\
\hline
3 & 10\ \ 3\ \ 0.98\ \ 0.05\ \ 5\ \ 4\  \     & 95,37\%  & 44\%  & 15\% \\
\hline
1 & $1000\ \ 0\quad \ 1\quad \ 0\quad \ \ 5\ \ 2\ \ \ \ $             & 100,00\% & 100\% & 0\%  \\
\hline
1 & $1000\ \ 3\quad \ 1\quad \ 0\quad \ \ 5\ \ 2\ \ \ \ $            & 100,00\% & 100\% & 0\% \\

\end{tabular}

\label{Table2}
\end{table}

\begin{table}[H]
\centering
\caption{Multiple-sample Test Hypotheses (Cut)}
\begin{tabular}{c|c|c|c|c}
\hline
 Label & $J_k$\ \ $A_k$\ \ $\theta$\ \ $\gamma$\ \ K\ \ $\Phi$ & Mean     & Min   & Std \\
\hline\hline
1 & $2\ \ 0\quad \ 1\quad \ 0\quad \ \ 5\ \ 2$             & 88,81\% & 47\% & 19\% \\
\hline
2 & 2\ \ 0\ \  0.98\ \ 0\ \ \ \ \ 5\ \ 2      & 89,27\% & 45\% & 21\% \\
\hline
3 & 2\ \ 0\ \ 0.95\ \ 0.01\ \ 5\ \ 2       & 89,50\% & 47\% & 20\% \\
\hline
3 & 2\ \ 0\ \ 0.98\ \ 0.05\ \ 5\ \ 2       & 90,27\% & 46\% & 20\% \\
\hline
1 & $2\ \ 3\quad \ 1\quad \ 0\quad \ \ 5\ \ 2$             & 97,53\% & 63\% & 9\%  \\
\hline
3 & 2\ \ 3\ \ 0.98\ \ 0.05\ \ 5\ \ 2       & 96,97\% & 63\% & 10\% \\
\hline
3 & 2\ \ 3\ \ 0.02\ \ 0.05\ \ 5\ \ 2       & 97,92\% & 63\% & 8\%  \\
\hline
1 & $5\ \ 0\quad \ 1\quad \ 0\quad \ \ 5\ \ 2$             & 86,02\% & 43\% & 21\% \\
\hline
2 & 5\ \ 0\ \ 0.02\ \ 0\ \ \ \ \ 5\ \ 2       & 87,16\% & 37\% & 22\% \\
\hline
1 & $5\ \ 3\quad \ 1\quad \ 0\quad \ \ 5\ \ 2$            & 93,73\% & 44\% & 16\% \\
\hline
2 & 5\ \ 3\ \ 0.02\ \ 0\ \ \ \ \ 5\ \ 2       & 94,59\% & 47\% & 15\% \\
\hline
3 & 5\ \ 3\ \ 0.02\ \ 0.05\ \ 5\ \ 2       & 96,27\% & 50\% & 13\% \\
\hline
1 & $10\ \ 0\quad \ 1\quad \ 0\quad \ \ 5\ \ 2\ $            & 82,85\% & 42\% & 23\% \\
\hline
1 & $10\ \ 3\quad \ 1\quad \ 0\quad \ \ 5\ \ 2\ $            & 92,36\% & 40\% & 17\% \\
\hline
2 & 10\ \ 3\ \ \ 0.02\ \ \ 0\ \ \ \ 5\ \ 2\  \     & 96,24\% & 50\% & 13\% \\
\hline 
3 & 10\ \ 3\ \ 0.02\ \ 0.05\ \ 5\ \ 2\  \     & 96,24\% & 50\% & 13\% \\

\end{tabular}

\label{Table3}
\end{table}

Due to limited space, we have included enough representative results to verify all the hypotheses and empirically show geometric test consistency. In Tables \ref{Table1}, \ref{Table2} and \ref{Table3}, we see test consistency with performance increasing with $J_k$ and $A_k$, and reaching 100\% for $J_k$ between 100 and 1000. The hypothesis that QHD+QCD improves QHD is verified (label $2 > 1$) for a range of parameters. The same applies for the hypothesis that the previous time-evolution of correlation improves QHD and QHD+QCD performances (label $3 > 2 > 1$). In Table \ref{Table3}, we see that all hypotheses are verified for multiple sampling with a fixed model configuration. Distance metric parameters were not scaled but could be which explains values close to 0 and 1 ($>100$ orders of magnitude). The best performance is for $\Phi=4$, with a less restrictive threshold for the rolling correlation network closeness in DMD. Results are not sensitive to the number of clusters, and we show them for K=5. Different asset classes have different market opening and closing hours that could have an impact on our analysis. For that, we use a cut version in Table \ref{Table2} and \ref{Table3}, with no missing values, and the uncut version with the full dataset in Table \ref{Table1}. Performance patterns are the same, but calibration parameters change. Finally, the clustering algorithm used is Kmedoids.

\section{Conclusion}

We can conclude that it is possible to improve the QHD in correlation structures for non-parametric, high-dimensional settings by combining it with QCD in a geometric version of the test via clustering. Also, we can improve both, QHD and QHD+QCD, by adding the pre-change time-evolution of correlation to the test with the geometric solution and solving the state-of-the-art issue of uncorrelated pre-change variables. Our geometric version of the test opens new doors by allowing us to include new sources of information. We have proved and empirically verified consistency and test hypothesis designs for the geometric test via clustering. This geometric solution can be applied to other statistical tests with similar properties such that consistency based on lemmas \ref{lemma1}, \ref{lemma2} and \ref{lemma3} can be proved. DMD is a new way to represent time-evolution of correlation respect individual asset variables. It is a useful measure of risk diversification for portfolios to avoid crowded risk positions. We verified that DMD improves QHD and QHD+QCD performances.\par

For future work it would be interesting to analyze the asymptotic high-dimensional ($p>>n$) properties of the DMD to come up with local summary statistics like in \cite{Banerjee2016} that could allow for a geometric solution based entirely on high-dimensional properties.

\section*{Acknowledgment}
The authors would like to thank Miralta Bank for the financial data, cloud services and discussions.

\bibliographystyle{IEEEtran}
\bibliography{main}

\begin{thebibliography}{10}
\providecommand{\url}[1]{#1}
\csname url@samestyle\endcsname
\providecommand{\newblock}{\relax}
\providecommand{\bibinfo}[2]{#2}
\providecommand{\BIBentrySTDinterwordspacing}{\spaceskip=0pt\relax}
\providecommand{\BIBentryALTinterwordstretchfactor}{4}
\providecommand{\BIBentryALTinterwordspacing}{\spaceskip=\fontdimen2\font plus
\BIBentryALTinterwordstretchfactor\fontdimen3\font minus
  \fontdimen4\font\relax}
\providecommand{\BIBforeignlanguage}[2]{{%
\expandafter\ifx\csname l@#1\endcsname\relax
\typeout{** WARNING: IEEEtran.bst: No hyphenation pattern has been}%
\typeout{** loaded for the language `#1'. Using the pattern for}%
\typeout{** the default language instead.}%
\else
\language=\csname l@#1\endcsname
\fi
#2}}
\providecommand{\BIBdecl}{\relax}
\BIBdecl

\bibitem{Markowitz1952}
H.~Markowitz, ``{PORTFOLIO} {SELECTION},'' \emph{The Journal of Finance},
  vol.~7, no.~1, pp. 77--91, mar 1952.

\bibitem{Junior2011}
L.~S. Junior and I.~D.~P. Franca, ``Correlation of financial markets in times
  of crisis,'' \emph{Physica A 391 (2012) 187--208}, 2011.

\bibitem{doi:10.1080/07474940801989202}
A.~G. Tartakovsky, ``Multidecision quickest change-point detection: Previous
  achievements and open problems,'' \emph{Sequential Analysis}, vol.~27, no.~2,
  pp. 201--231, 2008.

\bibitem{Banerjee2016}
T.~Banerjee and A.~O. Hero, ``Quickest hub discovery in correlation graphs,''
  in \emph{2016 50th Asilomar Conference on Signals, Systems and
  Computers}.\hskip 1em plus 0.5em minus 0.4em\relax IEEE, 2016, pp.
  1248--1255.

\bibitem{Banerjee2015}
T.~Banerjee, H.~Firouzi, and A.~O. Hero, ``Non-parametric quickest change
  detection for large scale random matrices,'' in \emph{2015 IEEE International
  Symposium on Information Theory (ISIT)}.\hskip 1em plus 0.5em minus
  0.4em\relax IEEE, 2015.

\bibitem{Hero2011}
A.~Hero and B.~Rajaratnam, ``Large-scale correlation screening,'' \emph{Journal
  of the American Statistical Association}, vol. 106, no. 496, pp. 1540--1552,
  2011.

\bibitem{Hero2012HubDI}
A.~O. Hero and B.~Rajaratnam, ``Hub discovery in partial correlation graphs,''
  \emph{IEEE Transactions on Information Theory}, vol.~58, pp. 6064--6078,
  2012.

\bibitem{Aminikhanghahi2017}
S.~Aminikhanghahi and D.~J. Cook, ``A survey of methods for time series change
  point detection,'' \emph{Knowledge and information systems}, vol.~51, no.~2,
  pp. 339--367, 2017.

\bibitem{article}
A.~Tartakovsky, ``Asymptotic properties of cusum and shiryaev's procedures for
  detecting a change in a nonhomogeneous gaussian process,'' \emph{Mathematical
  Methods of Statistics}, vol.~4, 04 1999.

\bibitem{Veeravalli2012}
V.~V. Veeravalli and T.~Banerjee, ``Quickest change detection,'' \emph{arXiv},
  2012.

\bibitem{Lorden1971}
G.~Lorden \emph{et~al.}, ``Procedures for reacting to a change in
  distribution,'' \emph{The Annals of Mathematical Statistics}, vol.~42, no.~6,
  pp. 1897--1908, 1971.

\end{thebibliography}
\vspace{10mm}

\appendices

\section{Proof of lemma \ref{lemma1}}
\begin{IEEEproof}

\begin{align}
    \left(\tau_G{\xrightarrow p}\left[\max_J{\left(J(i)\right)}\right]\right)|\ \forall\ i=1,\dots,m
\end{align}

\begin{align}
\begin{split}
\left(\tau_G{\xrightarrow p}\left[\max_i{f_V(V(i);J)}\right]\right)|\ J\rightarrow\ \max_J{(J(i))} 
\end{split}
\end{align}

\begin{align}
\begin{split}
\left(J{\xrightarrow p\theta|\ \theta=\max_i{\left(J(i)\right)}}\right)\Longleftrightarrow\left(\tau_G{\xrightarrow p}\left[\max_i{f_V(V(i);J)}\right]\right)
\end{split}
\end{align}

\begin{align}
\begin{split}
\left(J{\xrightarrow p\theta\ |\ \theta=\max_i{\left(J(i)\right)}}\right)\Longleftrightarrow\left(\tau_G{\xrightarrow p}\left[\max_i{\left(J(i)\right)}\right]\right)
\end{split}
\end{align}

\begin{align}
\begin{split}
\left(\max_i{\left(J(i)\right)}{\xrightarrow p}\max_i{f_V(V(i);J)}\right)|\ J\rightarrow\ \max_J{(J(i))},\\ 
\forall\ i=1,\dots,m
\end{split}
\end{align}

\end{IEEEproof}

\section{Proof of lemma \ref{lemma5}}
\begin{IEEEproof}
\underline{Categorical Distribution. RC subnetwork:}
\begin{align}
    X_1,\ldots\ldots\ {,X}_n\ \sim\ Cat\ (\theta),\   \
    P\left(X_i=j\ \right|\theta_j)
\end{align}
\begin{align}
    D=({rc}_1,\ \ldots,{rc}_n),\    \           {rc}_i=\left\{1,\ldots,n\right\}
\end{align}

With rc for rolling correlations. The probability of encountering a rc label (pair of variables) in a distance of rc through time from a distance matrix with threshold T, can be modeled by a categorical distribution:                

\begin{align}
\begin{split}
    P(D\ |\theta)=\prod_{i=1}^{n}{P\left({RC}_i=\ {rc}_i\ \right|\ \theta)=\ \prod_{i=1}^{n}{\theta_{rc_i}}}=\\
    \prod_{i=1}^{n}\prod_{j=1}^{n}\prod_{p=1 j\neq p}^{n}{(d_{jp}\le T)}^{{I\left[rc_i=j\right]+I\left[rc_i=p\right]}}=\\
    \prod_{i=1}^{n}\prod_{k_1=1}^{n}\prod_{k_2=1 k_1\neq k_2}^{n}{(d_{k_1k_2}\le T)}^{{I\left[rc_i=k_1\right]+I\left[rc_i=k_2\right]}}=\\
    \prod_{i=1}^{n}\prod_{k={k_1,k_2} k_1\neq k_2}^{m}{(d_k\le T)}^{{I\left[rc_i=k_1\right]+I\left[rc_i=k_2\right]}}= \prod_{i=1}^{n}\prod_{k=1}^{n}{\theta_{k}}^{rc_k}
\end{split}
\end{align}

With d, the distance between rolling correlations (\ref{equation26}), rc the pair of variable names in each correlation, k the pair of rolling correlations names (4 variable names), and I indicator function.

\underline{Dirichlet Distribution. Assets subnetwork:}
\begin{align}
\begin{split}
  \theta\ \sim\ Dir\ (\alpha),\\ 
  P\left(\theta\ \middle|\alpha\right) \sim\ \prod_{j=1}^{n}{\theta_j^{\alpha_j-1}I\left(\sum_{j=1}^{n}{\theta_j=1, \forall\ \theta_i\ }\right)}  
\end{split}
\end{align}

The probability of an individual asset being part of the pair of variables in a rolling correlation, rc, from a pair of rc in the distance d (\ref{equation26}), part of the distance matrix with a threshold T, can be modeled by a Dirichlet distribution:
\begin{align}
\begin{split}
        P\left(\theta\ \middle| a\right)=P\left(d\le T\ \middle| a\right) \ \sim\ \prod_{i=1}^{n}\prod_{k=1}^{n}{\theta_k^{a_{k-1}}}=\\
   =\prod_{i=1}^{n}\prod_{k=k_1,k_2;k_{11} \neq k_{12};k_{21}\neq k_{22}}^{m}{(d_k\le T)}^A,\\
   A=I\left[a_i=k_{11}\right]+I\left[a_i=k_{12}\right]+I\left[a_i=k_{21}\right]+\\
   +I\left[a_i=k_{22}\right]-1
\end{split}
\end{align}

Letter a, is the individual asset variable name. With d and k as before, $k_1$ being the asset names of one rolling correlation, $k_2$ the other rc pair of name variables, $k_{11}$ the name of an asset variable that is part of the rolling correlation $k_1$, same logic for other variables and indexes. By Bayes Theorem: 

\begin{align}
\prod_{i=1}^{n}{Cat\left(rc\ \right|d\le T)\ Dir(d\le T\left|a\right)=Dir(d\le T|rc+a)\ \ }
\end{align}
we want to prove the following link, for distributions N1 (for subnetwork 1 of rolling correlations), N2 (subnetwork 2 of asset variables), and the compound network, as N1+N2:

\begin{align}
\begin{split}
N1 \ \sim\ Cat\left(rc\ \right|d\le T),\ \ N2\ \sim Dir(d\le T\left|a\right),\\
N1+N2\ \sim Dir(d\le T|rc+a)
\end{split}
\end{align}

\begin{align}
\begin{split}
P\left(a\ \middle| rc\right)\ =\int{P\left(a\ |d\le T,\ rc\right)\ P\left(d\le T|rc\right)d\left(d\le T\right)}=\\ 
\int{P\left(a\ |d\le T\right)\ P\left(d\le T|rc\right)d\left(d\le T\right)}=\\
\int{P\left(d\le T|a\right)\frac{P\left(a\right)}{P\left(d\le T\right)}
P\left(rc|d\le T\right)\frac{P\left(d\le T\right)}{P\left(rc\right)}d\left(d\le T\right)}\\
=\int{P(d\le T|\ a)\ \ P(rc\ |d\le T)\ \frac{P\left(a\right)}{P\left(rc\right)}\ d\left(d\le T\right)}=\\ =\prod_{i=1}^{n}\prod_{k=k1,k2;k11 \neq k12;k21\neq k22}^{m}{(d_k\le T)}^A\times\\
\times\prod_{k_1=1}^{n}\prod_{k_2=1, k_1\neq k_2}^{n}{(d_{k_1k_2}\le T)}^B=\\
=Dir(d\le T|rc+a)\frac{P\left(a\right)}{P\left(rc\right)},\\
A=I\left[a_i=k_{11}\right]+I\left[a_i=k_{12}\right]+I\left[a_i=k_{21}\right]+\\
+I\left[a_i=k_{22}\right]-1,\\
B=I\left[rc_i=k_1\right]+I\left[rc_i=k_2\right]
\end{split}
\end{align}
\end{IEEEproof}

\end{document}